\documentclass[apj]{emulateapj}
\usepackage{natbib}
\usepackage{amsmath}
\def\be{\begin{eqnarray}}
\def\ee{\end{eqnarray}}
\usepackage{comment}
\usepackage{graphicx,subfigure}
\usepackage{appendix}

\expandafter\let\csname equation*\endcsname\relax
\expandafter\let\csname endequation*\endcsname\relax
\usepackage{amsmath,amssymb,amsfonts} 
\usepackage{graphicx} 
\usepackage{subfigure}
\newcommand{\dnud}{\Delta\nu_{\text{d}}}
\newcommand{\dtd}{\Delta t_{\text{d}}}

\newcommand{\bb}{\begin{bmatrix}}
\newcommand{\eb}{\end{bmatrix}} 

\newcommand{\gae}{\lower 2pt \hbox{$\, \buildrel {\scriptstyle >}\over {\scriptstyle
\sim}\,$}}
\newcommand{\lae}{\lower 2pt \hbox{$\, \buildrel {\scriptstyle <}\over {\scriptstyle
\sim}\,$}}
\newcommand{\us}{$\mu$s}

\submitted{Accepted to the Astrophysical Journal Letters 2014 July 1}
\shorttitle{LOFAR MSP Scintillation Study}
\shortauthors{Archibald, Kondratiev, Hessels, \& Stinebring}

\begin{document}
\title{Millisecond Pulsar Scintillation Studies with LOFAR: \\Initial Results} 
\author{Anne M. Archibald\altaffilmark{1}, Vladislav I. Kondratiev\altaffilmark{1,2}, Jason W. T. Hessels\altaffilmark{1,3}, 
\& Daniel R. Stinebring\altaffilmark{1,4}}
\affil{
$^{1}$ ASTRON, the Netherlands Institute for Radio Astronomy, Postbus 2, 7990 AA, Dwingeloo, The Netherlands\\
$^{2}$ Astro Space Center of the Lebedev Physical Institute, Profsoyuznaya str. 84/32, Moscow 117997, Russia \\
$^{3}$ Anton Pannekoek Institute for Astronomy, University of Amsterdam, Science Park 904, \\ 1098 XH Amsterdam, The Netherlands\\
$^{4}$ Oberlin College, Department of Physics and Astronomy, 110 North Professor Street, \\ Oberlin, OH 44074, USA\\
archibald@astron.nl, kondratiev@astron.nl, hessels@astron.nl, \& dan.stinebring@oberlin.edu
}

\begin{abstract}
High-precision timing of millisecond pulsars (MSPs) over years to decades 
is a promising technique for direct
detection of gravitational waves at nanohertz frequencies.
Time-variable, multi-path scattering in the interstellar medium is a significant source of noise for this detector, particularly as timing precision approaches 10~ns or better for MSPs in the pulsar timing array.
For many MSPs the scattering delay above 1 GHz is at the limit of detectability; therefore, we study it at lower frequencies.
Using the LOFAR 
(LOw-Frequency ARray)
 radio telescope we have analyzed short (5--20~min) observations of three MSPs in order to estimate the scattering delay at 110--190 MHz, where the number of scintles is large and,
hence, the statistical uncertainty in the scattering delay is small.
We used cyclic spectroscopy, still relatively novel in radio astronomy, on baseband-sampled data to achieve unprecedented frequency resolution while retaining adequate pulse phase resolution.
We detected scintillation structure in the spectra of the MSPs PSR~B1257+12, PSR~J1810+1744, and PSR~J2317+1439 with diffractive bandwidths  of 
$6\pm 3$, $2.0\pm 0.3$, and $\sim7$~kHz, respectively,  where the estimate for PSR~J2317+1439 is reliable to about a factor of 2.
For the brightest of the three pulsars, PSR~J1810+1744, we found that the
diffractive bandwidth has a power-law behavior $\dnud \propto \nu^{\alpha}$, where
$\nu$ is the observing frequency and $\alpha = 4.5\pm0.5$, 
consistent with a  Kolmogorov inhomogeneity spectrum.
We conclude that this technique holds promise for monitoring the scattering delay
of MSPs with LOFAR and other high-sensitivity, low-frequency
arrays like  SKA-Low.
\end{abstract}

\section{Introduction}
The effort to detect gravitational waves with an array of millisecond pulsars (MSPs) continues
to gain momentum.
One example of the maturity of this effort is the special focus issue of 
{\em Classical and Quantum Gravity} \citep{bjpw13}, comprised
of 16 articles on Pulsar Timing Arrays (PTAs) that detail their promise,
current status, and 
major challenges.
One substantial challenge is correcting for time-varying propagation delays due
to passage of the radio waves through the partially ionized interstellar medium
\citep[ISM;][]{sti13}.
The motion of the pulsar, ISM, and the Earth all contribute to the time variability
of various propagation delays, with different weighting  for each type
of delay.  
Frequency-dependent ($\propto \nu^{-2}$) dispersion, quantified by the dispersion measure (DM), 
produces
the largest time delay of typically tens to hundreds of ms at 1~GHz, depending
on the pulsar.
Because this is such a large effect compared to the $\sim$~10~ns timing correction
goal that is commonly pursued, all three  PTAs --- the North American Nanohertz Observatory
for Gravitational Waves, 
the European Pulsar Timing Array, 
and the Parkes Pulsar Timing Array ---
employ active DM-variability mitigation schemes \citep{dfg+13,lbj+14,kcs+13} .

The second-most-important delay, that due to multi-path scattering,  is generally not corrected for in timing efforts at frequencies near 1 GHz. The approximately $\nu^{-4}$ dependence of this delay makes it less important as a direct effect. For small scattering time delays $\tau \ll W$, where $W$ is the pulse width, the main effect is to delay the arrival of the pulse by $\tau$, with negligible time-smearing \citep{hs08, crg+10}.
Nonetheless, as we move towards timing pulsars at the 10-ns level of precision, it is important to begin estimating and correcting for this delay. One should also consider indirect effects of scattering delay on precision pulsar timing. For example, DMs to  pulsars included in the PTAs are determined by multiple-frequency observations, including frequencies near 400~MHz, where scattering delays can be substantial (about 150 times larger than at 1.4~GHz), with the potential to bias and cause time-variability in DM estimates and, hence, in the reported arrival time.

Here we present a pilot study for temporal monitoring of this scattering delay.
Although the frequencies of choice for high-precision timing of MSPs remain around 1--3 GHz,
we explore the scattering delay of MSPs at frequencies around 150~MHz for two reasons. 
First, the $\nu^{-4}$ dependence of $\tau$ means
that the effects of scattering are strong at these frequencies.
Second, the LOFAR {(LOw-Frequency ARray)} telescope provides 80~MHz of bandwidth
at these frequencies, where 1--2~MHz was typical for previous instruments.
 This, coupled with LOFAR's high sensitivity and versatility 
 \citep{hwg+13}, makes it an excellent instrument with which to explore scattering delays.
 As we discuss further below, the advent of a new signal processing technique --
 cyclic spectroscopy (CS) -- plays a central role in the results presented here.
 Prior to the availability of CS in radio astronomy, it was not possible to achieve the
 required fine frequency resolution and pulse phase resolution simultaneously.
 
In the past, several approaches have been taken to  estimate  the scattering
delay, $\tau$. 
These can, in general, be divided into time-domain
and frequency-domain  techniques,
depending on whether time-domain fitting of a scatter-broadening function
was applied to average profiles  \citep{bcc+04,lkm+01} or whether
an autocorrelation analysis or similar was applied to the scintillation-modulated
spectrum of the pulsar \citep{cwb85,kpss98,kpsk01,hs08}.
Although the correspondence between the scattering timescale and the
diffractive scintillation bandwidth (or decorrelation bandwidth), $\dnud$, is known
to be more complicated than this \citep{cr98}, the standard formula for
relating the two quantities is conventionally taken to be 
$2 \pi \tau \dnud = 1$, and we will use that in the present work.

Previous work has shown that time-domain fitting is complicated 
by the evolution of average
profiles with observing frequency \citep{hsw+13,pdr14}.
Although MSP profiles do not evolve with frequency as strongly
as do those of some slowly rotating pulsars, they typically have more
complex profiles with larger duty cycles. Both of those features
make time-domain fitting for scattering delay potentially inaccurate.
As an attractive alternative, we employ frequency-domain analysis of the 
spectra in order to characterize the diffractive bandwidth $\dnud$.

In \S2 we present the pilot observations  we have made for three MSPs.
Then, in \S3 we give details of our analysis procedure. 
Finally, in \S4 we present our results and discuss their
implications for the prospect of correcting MSP precision timing for
the effect of time-variable scattering delay.

\section{Observations and Initial Processing}
We used LOFAR baseband data from observations of \object[PSR B1257+12]{PSR~B1257+12}, \object[PSR J1810+1744]{PSR~J1810+1744}, and \object[PSR J2317+1439]{PSR~J2317+1439} obtained
for other purposes \citep{khs+14} on three occasions from late 2012 through early 2014 for the analysis described below.
Since PSR~J1810+1744 is in a black-widow system  \citep{hrm+11}, we ensured that our observation (at orbital phases 0.78--0.80) was well away from the eclipse centered at orbital phase 0.25.
All data were acquired in coherent beam-formed mode using the central LOFAR Core stations.
As described in \cite{sha+11}, \cite{hwg+13}, and \cite{khs+14}, baseband data were recorded and stored in 400 subbands of width 195 kHz spanning the frequency range from approximately 110--190~MHz.
Each 195-kHz subband was coherently dedispersed and then, in a manner described below, divided into 1024 frequency channels for a final
frequency resolution of $\Delta f = 190$~Hz spanning a total of 80~MHz.
This narrow frequency resolution was needed to analyze the kHz-scale 
modulation produced by multi-path scattering in this frequency range.

\begin{figure}
\begin{center}
        \includegraphics[width=3.25in, angle=0]   
        {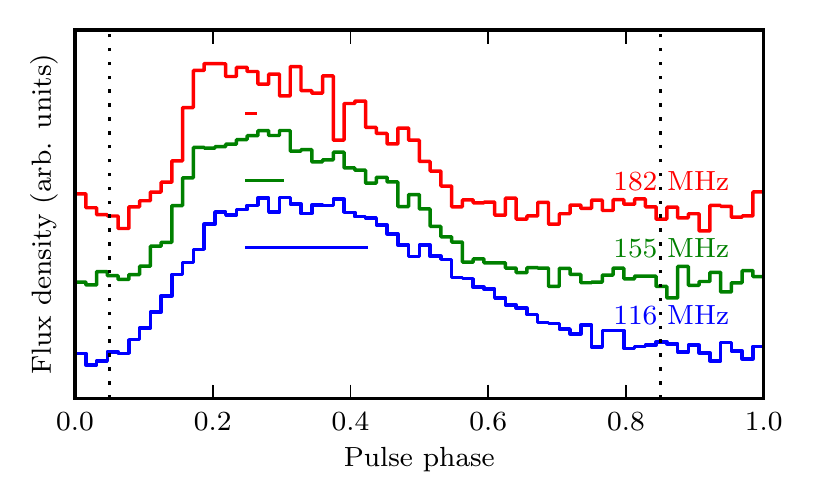} 
   \caption{Pulse profiles from our observation of PSR~J1810+1744. Each profile is based on 4~MHz of our 300-s observation. The dotted lines indicate the beginning and end of the region we have treated as on-pulse; all other regions are treated as off-pulse. Note the appearance of a tail at low frequencies, presumably due to scattering, and the resulting lack of a clear off-pulse region to use for bandpass and RFI subtraction. Horizontal bars indicate the scattering time $
\tau$ computed from our measured $\dnud$.
    	\label{fig:profiles}}
\end{center}
\end{figure}
Traditional pulsar spectral processing operates on a coherently dedispersed baseband data stream and Fourier-transforms\footnote{Some modern pulsar backends use polyphase filterbanks instead, obtaining better channel isolation at the cost of increasing the minimum product $\Delta f\Delta t$ by a factor of 4, 8, 16, or more.} short segments of duration $\Delta t$ to form an $N_f$-channel filter bank (with channels of width $\Delta f$). These spectra are then averaged modulo the pulsar period to form a (radio-frequency, pulse-phase-bin) 2D array that is accumulated for a subintegration length which is typically 10--60~s. This array then allows RFI excision and averaging over frequency to produce profiles like those in Figure~\ref{fig:profiles}, which are the primary input to pulsar timing. However, the constraint $\Delta f \Delta t \geq 1$ inherent in this approach poses a fundamental problem for scintillation studies of MSPs at low radio frequencies because small values of $\Delta f$ and $\Delta t$ are  needed simultaneously (e.g.\ we needed $\Delta f = 200$~Hz and $\Delta t = 50\;\mu$s for one of the pulsars in our study, requiring $\Delta f \Delta t = 0.01$). Fortunately, the application of CS to observations of pulsars \citep{dem11,wds13} makes such analysis  possible.
CS generalizes spectral analysis to  {\em periodically} varying signals \citep{ant07}. 
Therefore, one can choose
$\Delta t$ to be as long as a subintegration length (often tens of seconds) without impacting pulse phase resolution $\Delta t_{\phi}$. 
This separation of $\Delta t$ and $\Delta t_{\phi}$ offers the possibility to study scintillation in the frequency domain while still retaining $\Delta t_{\phi}$ short enough to resolve scattering tails in the pulse-phase domain. The computed quantity is known as the periodic spectrum \citep{dem11}, or in other application areas as the Wigner-Ville spectrum \citep{ant07}. Its interpretation ranges from straightforward, as  here, to
quite subtle, depending on the application.

In our processing, for each 195-kHz subband, periodic spectrum estimates were formed modulo the pulse period with a pulse-phase resolution $\Delta t_{\phi} = P/64$, where $P$ is the pulsar period. This resulted in $\Delta t_{\phi} \approx$ 100~$\mu$s, 25~$\mu$s, and 50~$\mu$s
for the pulsars PSR~B1257+12, PSR~J1810+1744,
and PSR~J2317+1439, respectively. 
The periodic spectrum is a real-valued second-order product of the baseband data that can be accumulated over time. It has dimensions $N_{\rm f}$ by  $N_{\rm \phi}$,
where $N_{\rm f}$ and  $N_{\rm \phi}$ are the number of frequency channels and pulse phase bins, respectively. 
We integrated the periodic spectrum for $\Delta t = 10$~s (shorter than the diffractive timescale) to form a set of $N_{\rm s}$ subintegrations producing a final data cube  denoted $C\;(N_{\rm f},\; N_{\rm \phi},\; N_{\rm s})$. 

This processing was accomplished with the
{\tt\small dspsr} program\footnote{\tt\small http://dspsr.sourceforge.net} \citep{sb11}, which incorporates
 CS as well as a wide range of state-of-the-art pulsar
 signal processing code.
 The  version of {\tt\small dspsr -cyclic} initially available to us allowed serious
 spectral leakage of narrow-band 
 radio-frequency-interference (RFI) signals.
Since the majority of interfering signals in the LOFAR high band are of this 
character\footnote{Offringa et al. 2013 \nocite{obz+13}found an average RFI occupancy of 3.2\%,
with most signals narrower than the 0.76-kHz resolution they employed.} 
we added a polyphase-filter-like step (activated with the {\tt\small -cyclicoversample} option) to the {\tt\small dspsr} code
and propagated this to the main code repository.
This sufficiently isolated narrow-band RFI, which was then removed in the next processing step.

\section{Analysis and Results}
Processing continued on 4-MHz (PSR~J1810+1744) or 8-MHz ``parts,'' which are the subband aggregates written by the beamformer; for each we assembled a data cube by concatenating the $N_{\rm b}$ data cubes for each subband in the frequency direction.
The real-valued periodic spectra (see, e.g., Figure~2 in Demorest 2011) are pulse-phase-resolved representations of the frequency structure imposed by multi-path scattering, but on a scale finer than $\Delta f \Delta t_{\phi} = 1$. It is easily verified from the construction of the periodic spectrum that averaging it over all phases exactly recovers the traditional spectrum. Omitting from this average phase ranges where there is no pulsar signal, either intrinsic or scattered, cannot affect the obtained spectrum, and allows estimation of the non-pulsar components of the measured spectrum.
This posed no problem for  processing the PSR~B1257+12 and PSR~J2317+1439 data, but it did cause some problems for analyzing the PSR~J1810+1744 data in the lower frequency range, as we comment on below.
\begin{figure}
\begin{center}
        \includegraphics[width=3.25in, angle=0]   
        {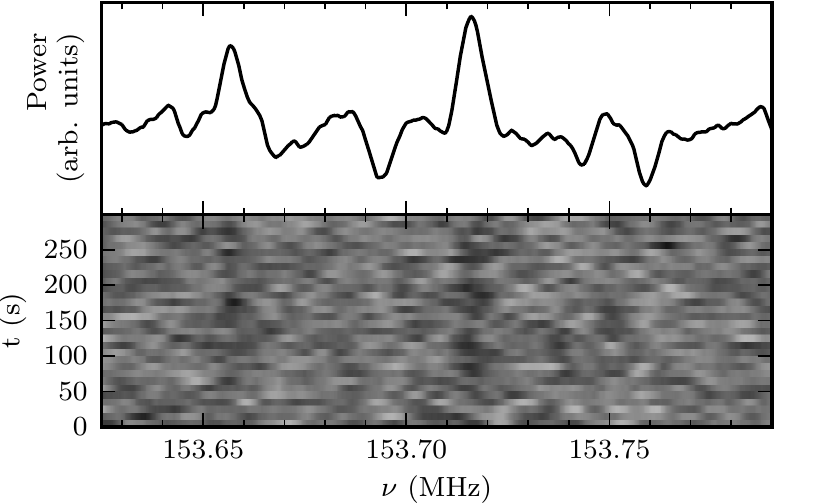} 
   \caption{Dynamic spectrum of a small frequency range for PSR~J1810+1744. The lower panel shows power as a function of radio frequency and time (darker indicating stronger), smoothed by $\dnud = 2$~kHz. The upper panel shows the result of averaging the same dynamic spectrum over time. Note the two bright scintles visible in this plot. Most scintles, even for this relatively bright pulsar, are too faint to see individually, appearing only when statistically combined using an autocorrelation.
    	\label{fig:dynspec}}
\end{center}
\end{figure}

For each pulsar, we chose an on-pulse phase gate of width $W_{\rm on}$ and a similar off-pulse gate of width $W_{\rm off} = P - W_{\rm on}$. By averaging over these phase ranges we created on- and off-pulse 2D data slices $D_{\rm on, off}\; (N_{\rm b}N_{\rm f},\; N_{\rm s})$.
We then formed the dynamic spectrum array $P (N_{\rm b}N_{\rm f},\; N_{\rm s}) = D_{\rm on} (N_{\rm b}N_{\rm f},\; N_{\rm s})\; -\; D_{\rm off} (N_{\rm b}N_{\rm f},\; N_{\rm s})$.
Requiring $W_{\rm on} > W + \tau$, where $W$ is the intrinsic pulse width and $\tau$ is the duration of the multi-path scattering tail, ensures that this recovers the traditional dynamic spectrum.
We show a very small section of the dynamic spectrum for PSR~J1810+1744 in Figure~\ref{fig:dynspec}, where we focus attention on several bright interference maxima or ``scintles''.

Although RFI had been isolated by the polyphase filtering step, it was still a significant problem for our frequency-domain analysis.
We manually excised all regions in frequency-subintegration space containing RFI spikes.
We also corrected for the digitally-determined bandpass shape of each 195-kHz subband imposed by a front-end polyphase filter and blanked out the edges of each polyphase subband, where the power aliased from the other side of the subband was more than 10\% of the total power; this required blanking approximately 10\% of the band in every observation.

\begin{figure}
\begin{center}
        \includegraphics[width=3.25in, angle=0]   
        {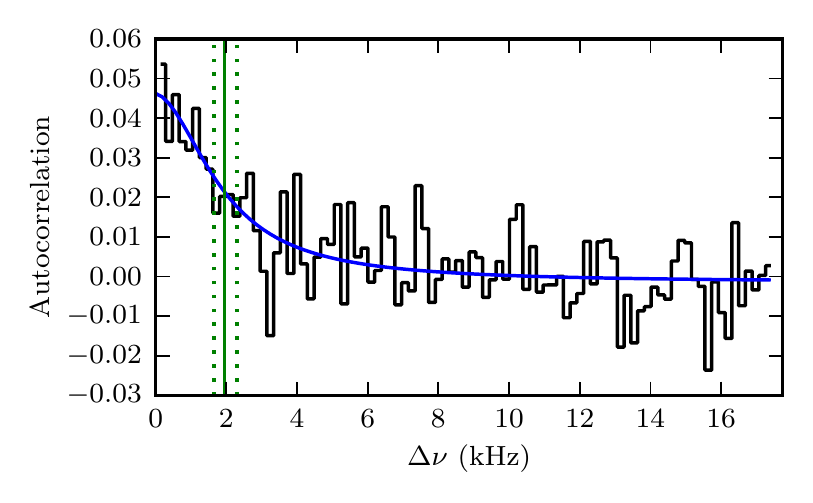} 
   \caption{Autocorrelation function (ACF) for the spectrum of PSR~J1810+1744 in the 4-MHz frequency band centered on 155~MHz. The total observation time is 300~s. The black curve is the ACF at a resolution of 190 Hz. The blue line is the best-fit Lorentzian (see text), which has a HWHM marked by the green vertical line of $\dnud = 2.0 \pm 0.3$~kHz, corresponding to a scattering time delay of  $\tau = 80 \pm 10$~\us.
    	\label{fig:acf}}
\end{center}
\end{figure}

Once the RFI excision was accomplished, we  analyzed each 4- or 8-MHz part individually using a standard autocorrelation function (ACF) analysis (e.g. Cordes et al.\ 1985) on the final spectra comprised of either 20480  channels (for PSR~J1810+1744) or 40960 channels. \nocite{cwb85}
The resulting ACF was then fit to a Lorentzian function
$\rho_S (\delta\nu) = A + {B}/[1+(\delta\nu / \dnud)^2]$,
where $\dnud$ is the conventional diffractive bandwidth
(HWHM of the frequency ACF function for small values of $A$, as was the case
here), where the offset $A$ and scale-factor $B$
are not relevant to this study.
We ignore the zero-lag noise spike as part of this fit, but use it to estimate the signal-to-noise ratio $R_\rho$.
A Lorentzian ACF is the appropriate functional form if
the image point-spread function due to scattering is Gaussian and, hence, the pulse-broadening function is a one-sided exponential \citep{ric90}.
To aid manual inspection of fit quality, we downsampled the dynamic spectrum where necessary to obtain approximately 16 bins across the ACF peak.
An example of calculated ACF and fitted function for PSR~J1810+1744 are
shown in Figure~\ref{fig:acf}.

A simple calculation of the signal-to-noise (S/N) of the ACF yields the expression
\be
R_\rho = b\;  \left(\frac{S_{\rm avg}}{\rm SEFD}\right)^2  \sqrt{\dnud \dtd B\; T}.
\label{eqn:snr}
\ee
Here,  $S_{\rm avg}$ is the phase-averaged pulsar flux density, SEFD is the system equivalent flux density,
$B$ is the total bandwidth, and $b$ is a dimensionless constant of order unity.
This assumes that the spectra are integrated for the diffractive scintillation time $\dtd$; after that,
the ACFs are incoherently averaged up to the total integration time $T$.
Raw sensitivity is clearly vital because of the squared ratio of signal strength to system noise.
Note that because $\dnud \propto \nu^4$ and $\dtd \propto \nu$ for anticipated ISM conditions,
we expect that $R_\rho \propto \nu^{5/2}$, which should be visible across the LOFAR
band, although the SEFD 
and gain degrade away from the center of the band \citep{hwg+13},
which will partially offset this improvement.

A summary of relevant observational details and the results of this ACF fitting process are presented in Table~\ref{table:parms}. 
As indicated there, the fitted parameters for PSR~J1810+1744 are well constrained.
However, the ACFs for the other two pulsars were much weaker, as indicated by the signal-to-noise ratio $R_\rho$ and by the error estimates on $\dnud$ in Table~\ref{table:parms}.
This follows because the S/N of the time-domain data (e.g.\ the time-averaged pulsar flux divided by the noise rms in the pulse profile, 
corrected for the number of phase bins) was a factor of 3--4 smaller than for PSR~J1810+1744, and Equation (\ref{eqn:snr}) shows that this will result in a factor of 9--15 degradation in $R_\rho$ if the SEFD is comparable for each of these pulsars.
\begin{deluxetable*}{cclcccc}
\tabletypesize{\scriptsize}
\tablecolumns{7}
\tablecaption{Scintillation parameters}
\tablehead{
\colhead{PSR} & \colhead{MJD} &  \colhead{ObsID} & \colhead{Duration} & \colhead{DM} & \colhead{$\dnud$}   & \colhead{$R_\rho$}  \\
 & & & \colhead{(s)} & \colhead{(${\rm pc\;cm^{-3}}$)} & \colhead{(kHz)}  }
\startdata
B1257+12 &  56280 & L81046 & 1200 & 10.2 & $6\pm 3$ & 7.9 \\ 
J1810+1744 &  56693 & L203594 & 300 & 39.7 & $2.0\pm 0.3$ & 21.0 \\ 
J2317+1439 &  56286 & L81273 & 1200 & 21.9 & $\sim 7$ & 4.1 
\enddata
\label{table:parms}
\tablecomments{The value for $\dnud$ is referenced to an observing frequency of 
$\nu = 150$~MHz. The value of DM quoted here is for comparison purposes only. $R_\rho$ here is the peak value of the ACF divided by the noise averaged over a bandwidth of $\dnud$. Note than in addition to their longer duration, the fits quoted here for pulsars PSR~B1257+12 and PSR~J2317+1439 used 8 MHz of bandwidth rather than 4 MHz.}
\end{deluxetable*}

\begin{figure}
\begin{center}
        \includegraphics[width=3.25in, angle=0]   
        {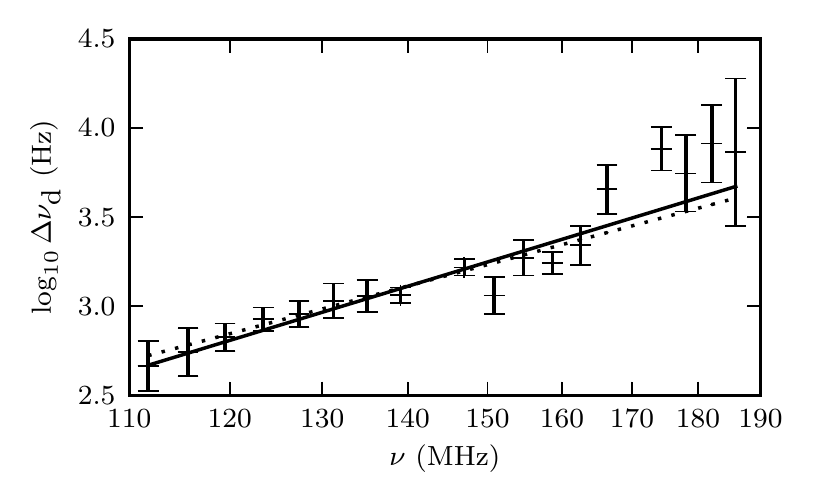} 
   \caption{Diffractive bandwidth $\dnud$ as a function of frequency for PSR~J1810+1744 in the LOFAR high band. The best-fit power law is shown as a solid line and the best-fit power law with a slope of 4 as a dotted line.
    	\label{fig:dbw}}
\end{center}
\end{figure}

The behavior of diffractive scintillation bandwidth $\dnud$ as a function of observing frequency is of great interest because it is influenced by the distribution of scattering material along the line of sight, the nature of the inhomogeneity spectrum, and the transverse extent of scattering ``screens'' (see Cordes and Lazio 2001 and references therein).\nocite{cl01}
In Figure~\ref{fig:dbw} we show results for the three MSPs in this pilot study. 
Only the PSR~J1810+1744 data are of high enough quality to comment on the log~$\dnud$ vs.\ log~$\nu$ slope over the LOFAR band. 
We find a logarithmic slope of $\alpha = 4.5 \pm 0.5$ over this range, consistent with predictions for a thin-screen, Kolmogorov turbulence model of unlimited transverse extent ($\alpha = 4.4$), but also consistent with numerous other plausible models \citep{ric90,lr00,cl01,lkm+01,bcc+04}.
In particular, a break in the power-law to a smaller value of $\alpha$ at lower frequencies could indicate
an ``inner scale'' to the density variations or a truncation of the scattering disc at large spatial scales \citep{ric90,cl01}.

We note that for the PSR~J1810+1744 data selection of an on-pulse region poses a difficult decision. 
Including phase ranges where there is little or no signal reduces the signal-to-noise. 
However, while averaging the periodic spectrum over all signal-containing phases does produce the familiar spectrum, examining both simulated spectra and those observed for PSR~B1937+21 \citep{dem11} shows that the frequency structure of a scattered pulse narrows as one moves to later phases in which the scattering tail dominates. 
Omitting these later phases biases the $\dnud$ estimate upwards. 
Since the scattering tail lengthens substantially at lower frequencies, 
this could produce a break in the power law relation between $\dnud$ and $\nu$,
mimicking the effect of an inner scale or a truncated screen.

\section{Discussion and Conclusions}
This effort aimed at exploring  LOFAR's potential for frequency-domain studies of multi-path scattering in the ISM.
Based on results from the bright MSP PSR~J1810+1744 and two other moderately bright MSPs we find good potential for further studies, which we are embarking upon.
We have shown that CS is a powerful and essential tool for studying MSP scintillation at LOFAR frequencies, and we have improved upon its implementation in the standard pulsar signal-processing package {\tt\small dspsr}.
This pilot study also serves to
  demonstrate the power that the low-frequency (50-350~MHz) component
  of the Square Kilometre Array (SKA-Low) will have for such studies.
  SKA-Low will provide an order-of-magnitude improvement in
  sensitivity over LOFAR, and can thus serve as a powerful ISM monitor
  to support high-precision timing at higher observing frequencies.

There are two major challenges to expanding these studies to other MSPs. 
The first problem is raw sensitivity. Despite LOFAR's large collecting area and
state-of-the-art architecture, we only had borderline detections of 
scintillation structure for two 
 moderately bright MSPs.
 Admittedly this was in 1200-s data blocks --- and we note that these observations
were earlier in the commissioning process --- but Equation (\ref{eqn:snr}) emphasizes that 
narrow $\dnud$ requires high instantaneous sensitivity or a compensating
increase in (incoherent) integration time.
{Because of the lower sky temperature at higher frequencies, some MSPs will
be better studied in the range 300--500~MHz using this method.}
Secondly, as was true for PSR~J1810+1744 in this study, a combination of large duty cycle
($W/P$) and/or substantial values of $\tau$ at low frequency means that there
will be limited or no off-pulse baseline for some pulsars.
This makes it difficult to determine the traditional spectrum accurately
from the periodic spectrum, which possesses interference structure out to at least the
quadratic sum of $W$ and $\tau$.
However, in future observations of PSR~J1810+1744 and similar high-duty-cycle or heavily
scattered MSPs, we plan to use LOFAR's multi-beaming capability to provide off-source calibration. For very heavily scattered pulsars it may also be necessary to apply these techniques at higher frequencies, using baseband-output or realtime cyclic-spectroscopy backends of other telescopes.

The ultimate goal of this work is to improve the timing accuracy of MSPs at
gigahertz frequencies.
In order to accomplish that we plan to expand the sample of MSPs studied
with LOFAR and to make multi-epoch observations of them.
We will compare these results with what is known 
at frequencies around 1~GHz
about multi-path scattering
and its temporal variations  for many of these pulsars \citep{lev+14}.
As has become common for studies attempting
to improve the pulsar-based gravitational-wave detector, this work is
likely to lead to increased understanding of another
area of astrophysics, in this case the ionized ISM.

This work was made possible by an ERC Starting Grant (DRAGNET) and NWO Vrije Competitie grant to J.W.T.H., and an NWO Bezoekersbeurs to support the long-term visit of D.S.
Other important  support includes NSF awards 0968296 (PIRE) and 1009580 (RUI) as well as a Research Status award from Oberlin College to D.S.
We thank A.~Karastergiou, B.~Stappers, and J.~Verbiest for useful discussions and other contributions.
The data were taken under proposals LC0-011 and LC1-027 (PI: Verbiest). We thank the referee for numerous helpful suggestions.

% REINSTATE if references change!  \bibliography{../journals_apj,../scintillation,../modrefs,../psrrefs,../crossrefs}

\end{document}